\documentclass[11pt]{article}
\usepackage[margin=1in]{geometry}
\usepackage{amsmath, amssymb}
\usepackage{booktabs}
\usepackage{graphicx}
\graphicspath{{figures/}}
\usepackage{xcolor}
\usepackage[colorlinks=true, allcolors=blue]{hyperref}
\usepackage{enumitem}
\usepackage{natbib}

\newcommand{\fbfa}{\text{FB5A}}

\title{\textbf{How the fly holds a single goal:\\
normalization, not selection, in \emph{Drosophila} FC2}}
\author{Gioele Nanni \\ Common Origin \and Christopher Lee \\ Common Origin}
\date{}

\begin{document}
\maketitle

\begin{abstract}
A walking fly steers toward a goal direction, held as a bump of activity across the FC2 neurons of the
fan-shaped body. These neurons also inhibit one another over distance, more
strongly the farther apart they are, a feedback proposed to keep the fly on a single goal. We asked, from the connectome, what circuit produces this inhibition,
and whether it lets FC2 actively \emph{choose} one goal among competitors (a winner-take-all) or simply
keeps a goal set elsewhere as one clean bump. Tracing the wiring in a single FlyWire brain,
we find the inhibition is almost entirely global: four \fbfa{}
cells inhibit every FC2 neuron roughly equally, with a smaller, distance-dependent contribution from
$h\Delta$ interneurons and a negligible direct component. A ring-attractor winner-take-all (the kind
the compass uses) requires local recurrent excitation that the FC2 wiring lacks, so this geometry
cannot build one; and across a range of dynamical models, including a spiking network, no version of
the circuit locks onto a winner at the connectome-scaled reference coupling. FC2 therefore \emph{normalizes} an externally set goal rather than
selecting it, with \fbfa{} likely acting as the global normalizer, much as the APL neuron does in the
mushroom body. We are explicit about two open points: a different mechanism, mutual
inhibition between two competing goals (which $h\Delta$ supplies), could in principle select at very
strong coupling, and we bound rather than exclude it; and \fbfa's inhibitory identity is a
low-confidence prediction of the connectome's transmitter classifier, not yet measured, and likely not
GABAergic. We then ask where the goal is actually set: the connectome nominates an
upstream $h\Delta$ network and rules out the leading proposed alternative, whose
neurons supply under $0.2\%$ of FC2's input. Finally, we propose a direct experiment, silencing
\fbfa{} while imaging FC2, that would test the account.
\end{abstract}

% ============================================================================
\section{Introduction}
\label{sec:intro}

Navigating animals hold a \emph{goal}: a direction, in world coordinates, they intend to travel.
In \emph{Drosophila}, menotaxis, holding a fixed bearing relative to a distant cue, is the
paradigmatic case \citep{giraldo2018, green2019}, and the central-complex circuit that implements it is now well characterized.
A heading estimate is maintained as a single bump of activity across the EPG ``compass'' neurons
\citep{seelig2015, kim2017}; a goal direction is maintained as a bump across the FC2 neurons of
the fan-shaped body \citep{mussellspires2024}; and the PFL3 neurons convert the heading--goal
difference into a left--right steering command, while PFL2 neurons, tuned to the anti-goal
direction, set the \emph{gain} of that command \citep{westeinde2024, mussellspires2024}.

\citet{mussellspires2024} also reported a suggestive property of the goal bump itself:
optogenetically activating FC2 in one fan-shaped-body column \emph{suppresses} FC2 activity in
other columns, and \emph{more strongly the farther away}, feedback inhibition they proposed
serves to keep a single goal bump. The mediating circuit and its computational character were left
open. That is our starting point. We ask, from the connectome: \emph{what implements this feedback
inhibition, and is it the kind of local lateral inhibition that would build a winner-take-all, or
something else?}

\paragraph{The intuitive reading, and why it needs testing.}
The natural reading of ``single-bump-enforcing inhibition'' is a winner-take-all like the EPG
compass, which selects one heading through a continuous ring attractor, \emph{local} recurrent
excitation between neighbouring columns, balanced by global inhibition \citep{kim2017}. The
connectome even offers a clean inhibitory candidate: \fbfa, four tangential cells, \emph{predicted}
GABAergic by the FlyWire transmitter classifier \citep{eckstein2024} (confidence $\sim$0.79; no experimentally verified
transmitter, so we treat the inhibitory identity as a prediction), that contact \emph{all} 85
FC2 neurons (7{,}741 synapses). But a WTA needs local excitation, and whether the FC2 circuit has
it is an empirical question the connectome can answer.

\paragraph{Contributions.}
(i) We \emph{decompose} the reported distant inhibition into a uniform (\fbfa) component, an
anti-local ($h\Delta$) component, and a negligible direct component (Sec.~\ref{sec:res-decomp}), 
sharpening \citet{mussellspires2024} and the known $h\Delta$ vector-summation motif
\citep{hulse2021, lyu2022} into a cell-type-resolved account. (ii) We show the inhibition is global and
anti-local, \emph{not} local, so the circuit cannot host a within-FC2 ring-attractor WTA, a
constraint we verify across model families with a bistability test, the wiring replicating in a
second connectome (Secs.~\ref{sec:res-decomp} and~\ref{sec:res-nowta}); this reconciles the ``single-bump'' inhibition of
\citet{mussellspires2024} with the fact that the FC2 wiring is not a ring attractor of the kind now
modelled for adjacent fan-shaped-body circuits \citep{lanz2025}. (iii) We interpret \fbfa{} as a
uniform normalizer, the APL motif implemented by a specific cell (Sec.~\ref{sec:res-normalizer}).
(iv) Because selection is upstream, we ask the connectome \emph{where}: it nominates the
$h\Delta$C-led recurrent network as the goal-holding substrate FC2 reads (valence-gated via FB5AB, a
distinct sibling cell type of \fbfa{}), and \emph{excludes} the Lanz $h\Delta$K--PFG attractor
as FC2's source, since it supplies $<$0.2\% of
FC2's input (Sec.~\ref{sec:res-selector}). (v) We derive a falsifiable prediction: under two-cue
competition, silencing \fbfa{} preserves the relative activation of the competing columns (a global
\fbfa) rather than changing which dominates (a local selector), with overall disinhibition as a
secondary signature (Sec.~\ref{sec:res-prediction}). We are explicit throughout about what is
earned (the decomposition, the no-local-WTA constraint, and the $h\Delta$K--PFG exclusion, all
transmitter-independent) versus offered (the normalizer role and the goal-substrate nomination,
falsifiable candidates we specify direct experiments to test).

% ============================================================================
\section{Related work}
\label{sec:related}

\paragraph{The heading--goal--steering axis.} The EPG ``compass'' maintains a single heading bump
through a continuous ring attractor, local recurrent excitation balanced by global inhibition
\citep{seelig2015, kim2017}, the canonical central-complex winner-take-all, and the architecture
we ask whether FC2 copies. Downstream, PFL3/PFL2 read the FC2 goal against the compass heading to
produce an opponent steering command \citep{westeinde2024, mussellspires2024}. \citet{mussellspires2024}
established the FC2 goal bump and reported the distant feedback inhibition this paper decomposes.

\paragraph{Fan-shaped-body circuitry and goal models.} \citet{hulse2021} reconstructed the
central-complex connectome, characterized the $h\Delta$ vector-summation motif (half-fan-shaped-body
offset), and framed the fan-shaped body as a context-dependent action-selection centre; \citet{lyu2022} then
demonstrated experimentally that the $h\Delta$B layer performs this vector summation in behaving flies.
\citet{matheson2022} showed distributed odour-and-wind cues are integrated within the fan-shaped body
(via $h\Delta$C and tangential gating) into a navigational goal. \citet{lanz2025} modelled a
fan-shaped-body goal circuit ($h\Delta$K$+$PFG) as a disinhibition-gated recurrent attractor read out
by PFL, the strongest persistent-goal mechanism proposed to date (currently a preprint), and the
alternative we test FC2 against.

\paragraph{The normalization motif and transmitter caution.} Global feedback inhibition that
normalizes a population into a sparse code is the motif the mushroom-body APL neuron implements
\citep{lin2014, flyhash2017}; APL inhibition is itself spatially graded rather than perfectly uniform
\citep{amin2020}. Transmitter identity in the fan-shaped body is treacherous: dorsal-FB / 23E10
tangential neurons long reported GABAergic were recently shown to express no GABA
\citep{jones2025}, and in the fly olfactory system glutamate is inhibitory via GluCl$\alpha$
chloride channels \citep{liuwilson2013}, with GABA receptors (Rdl/GABA-B) present in the central
complex \citep{enell2007}. These bracket the transmitter-contingent parts of our account.

% ============================================================================
\section{Methods}
\label{sec:methods}

\subsection{Connectome data and the preferred-bearing axis}
\label{sec:m-data}
We source all wiring from a single FlyWire brain \citep{dorkenwald2024, schlegel2024} so that every
connection is a real reconstructed edge rather than an assumed one; predicted transmitters are the
FlyWire connectome's synapse-image neurotransmitter classifier \citep{eckstein2024}, surfaced per neuron
by \citet{schlegel2024}. The relevant populations are the 85 FC2 neurons, the four
\fbfa{} neurons, and the $h\Delta$/$v\Delta$ interneurons. Each FC2 neuron has a preferred bearing
$\psi_i$ recovered from the spectral structure of the FC2/PFL columnar connectivity fingerprint (not
from \fbfa{} or $h\Delta$, so bearing-based tests on those cells are not circular). Route weights are
disynaptic products $\text{profile}(\,W_{\text{FC2}\to X}\,W_{X\to\text{FC2}})$ binned by bearing
distance, on a common synapse-count scale. All measurements are repeated in the independently
reconstructed hemibrain \citep{scheffer2020, hulse2021}, where the real anatomical column index gives
an axis independent of the spectral one.

\subsection{The bistability (seeded-basin) test and model families}
\label{sec:m-bistab}
Operationally, a winner-take-all is \emph{bistable}: which candidate wins depends on the initial
state. We drive a model with two competing goals, seed the dynamics toward the left vs.\ the right
candidate, and measure the \emph{basin gap} = (settled bump location $\mid$ seeded-right) $-$
(settled $\mid$ seeded-left). A gap $\approx 0^\circ$ means one input-determined fixed point (no
WTA); a large gap means history-dependent latching (a WTA). We test, each with connectome-imposed
connectivity and a handful of free gains: \textbf{(1)} subtractive inhibition
$x_i=[\,\text{drive}_i-\text{global inhibition}\,]_+$; \textbf{(2)} a spiking-proxy (noisy-rate)
model, and separately a committed leaky integrate-and-fire network (85 FC2 neurons with
threshold/reset/leak and \fbfa{} global-inhibition feedback); \textbf{(3)} divisive normalization
(Sec.~\ref{sec:m-model}); \textbf{(4)} the measured $\text{FC2}\!\to\!h\Delta\!\to\!\text{FC2}$
antipodal mutual-inhibition matrix; and \textbf{(5)} the one \emph{local} route, $v\Delta$, run as
local \emph{excitation} (the ingredient a ring-attractor WTA needs). Two controls use an idealized
uniform ring: a positive control adds local excitation (should latch), a negative control is
global-only (should not). We use a no-latch bound of $30^\circ$ and a WTA bound of $60^\circ$; the
positive control lands at $114^\circ$, validating the detector. To exclude non-recurrent selection we
additionally sweep the divisive sharpening exponent $p$ (feedforward max-selection).

\subsection{The divisive-normalization model}
\label{sec:m-model}
We instantiate the account as divisive normalization \citep{heeger1992, carandini2012}, a reusable
global-inhibition module on the real FC2 ring (a recurrent variant of the canonical form: the
sharpening exponent $p$ acts on the feedforward drive in the numerator, while the denominator pools the
recurrent population activity). Each FC2 neuron receives a drive; \fbfa{} feeds back
inhibition proportional to the pooled population activity, divided across the ring:
\begin{equation}
  x_i \;=\; \frac{\big[\,\text{drive}_i\,\big]_+^{\,p}}
                 {\sigma \;+\; g_{\text{inh}}\, w_i \,(\mathbf{pool}\cdot \mathbf{x})},
  \label{eq:divnorm}
\end{equation}
where $\text{drive}_i$ is the feedforward input to FC2 cell $i$ (a von Mises bump centred on the
goal bearing, concentration $\kappa$), $w_i$ are the real \fbfa$\to$FC2 per-cell inhibition weights,
$\mathbf{pool}$ the real FC2$\to$\fbfa{} pooling weights (both from FlyWire), and
$(p, g_{\text{inh}}, \sigma)$, with $\sigma$ the semi-saturation constant, are gains set the fly's way, from a task-agnostic objective (a single
clean bump for a single committed goal), \emph{never} from any downstream task. We report the spatial
non-uniformity of an inhibition profile as its \emph{modulation}, the peak-to-peak range across bearing
bins divided by the mean (so $0$ is perfectly uniform); the two-cue sweep parameter $\alpha$ scales an
imposed cosine non-uniformity that replaces the measured $w_i$ in the sweep only,
$w_i\!\propto\!1+\alpha\cos\psi_i$. Setting $g_{\text{inh}}=0$ silences \fbfa{} (the
prediction's control). We read two quantities from two integrations of Eq.~\ref{eq:divnorm}: a
per-step-normalized iteration for the bump's \emph{shape} (scale-free), and an un-normalized
iteration for the settled \emph{amplitude}. Reported concentrations are from the normalized form; the
printed equation iterated raw gives a matching $0.88$. The feedforward co-representation and two-cue
ratio results below use the shipped per-step-normalized form; because both are scale-invariant ratios,
they are unchanged by the normalization (we re-ran raw Eq.~\ref{eq:divnorm}: loser-mass and ratio move
by $<$1\%). The bistability test alone omits per-step normalization, which would otherwise clamp the
runaway a WTA needs. For robustness we sweep the free gains over a $3\times3\times3$ grid (power $\times\{0.7,1.0,1.4\}$,
$g_{\text{inh}}\times\{0.5,1,2\}$, $\sigma\times\{0.5,1,2\}$ about the committed operating point).

\subsection{The two-cue competition test}
\label{sec:m-comp}
To distinguish a spatially-uniform \fbfa{} from a within-FC2 local selector, we present two competing
goals (two drive bumps $120^\circ$ apart, amplitude ratio $1{:}0.7$) and compare the competing-column
activation ratio with \fbfa{} on vs.\ silenced, quantified as
$|\log(\text{ratio}_{\text{on}}/\text{ratio}_{\text{off}})|$. We repeat this for three uniform
functional forms (divisive, subtractive, excitatory-additive) and against a local-excitation selector
control, and sweep the spatial non-uniformity $\alpha$ of the inhibition.

\subsection{Input tracing and the selector nomination}
\label{sec:m-selector}
We classify FC2's presynaptic types (directional columnar $h\Delta$/PFN vs.\ broad FB-tangential),
scoring each by synapse mass and per-cell bearing concentration (circular resultant of a cell's
FC2 targets), and trace two-hop routes (MBON$\to$FB5AB$\to\{h\Delta$C, FC2$\}$; $h\Delta$K/PFG
in/out of FC2). Behaviour anchors are measured signatures from FC2 imaging \citep{mussellspires2024} plus one
flagged cross-system analogy \citep{toepfer2018}, not fits.

% ============================================================================
\section{Results}
\label{sec:results}

\subsection{The feedback inhibition decomposes into a uniform, an anti-local, and a negligible term}
\label{sec:res-decomp}

The reported distant inhibition could arise through three routes, direct
$\text{FC2}\!\to\!\text{FC2}$ synapses, the tangential \fbfa{} pool, or the $h\Delta$/$v\Delta$
interneurons, with very different computational meanings. We measure each.

\begin{figure}[t]\centering
\includegraphics[width=0.62\textwidth]{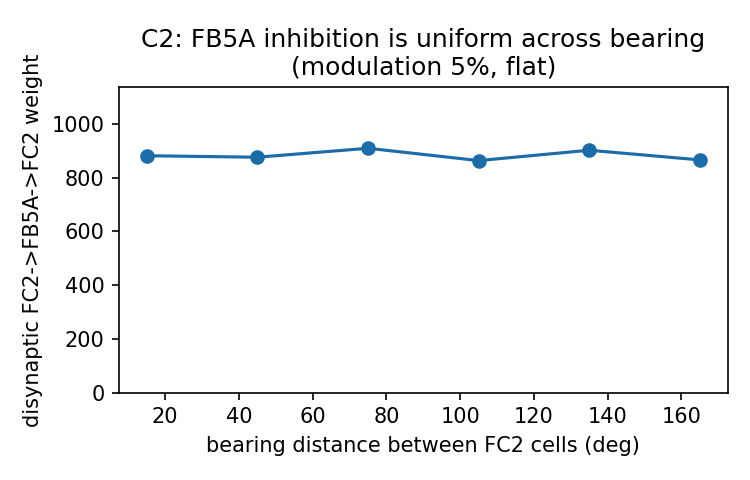}
\caption{\textbf{C2, \fbfa{} inhibition is uniform across bearing.} The disynaptic
$\text{FC2}\!\to\!\fbfa\!\to\!\text{FC2}$ loop weight is flat as a function of the bearing distance
between FC2 cells (modulation $5\%$): \fbfa{} is a global scaler, not a spatial selector. Measured
from the FlyWire connectome (\texttt{analyze\_fc2\_scoping.py}).}
\label{fig:c2}
\end{figure}

\paragraph{\fbfa{} is a global, not spatial, inhibitor.} If \fbfa{} were the inhibitory arm of a WTA,
its inhibition of an FC2 neuron would depend on that neuron's \emph{bearing distance} from the active
bump. It does not (Fig.~\ref{fig:c2}): the four \fbfa{} cells reach all 85 FC2 (85/85); the per-cell
\fbfa{}$\to$FC2 weight is uncorrelated with bearing (correlation with $\cos\psi$/$\sin\psi$ of
$+0.02$/$-0.04$); and the full disynaptic $\text{FC2}\!\to\!\fbfa\!\to\!\text{FC2}$ loop is
\emph{flat} across bearing distance (modulation $5\%$). \fbfa{} scales the whole population; it does
not sculpt it spatially.

\begin{figure}[t]\centering
\includegraphics[width=0.6\textwidth]{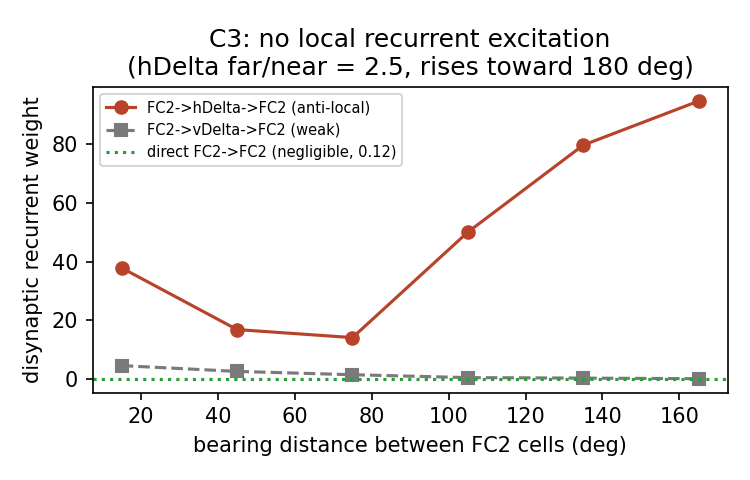}
\caption{\textbf{C3, no local recurrent excitation.} Direct $\text{FC2}\!\to\!\text{FC2}$ is
negligible; the $\text{FC2}\!\to\!h\Delta\!\to\!\text{FC2}$ disynaptic weight \emph{peaks at
antipodal bearings} (anti-local, near $37.8\to$ far $94.7$, far/near $2.5$, dipping mid-range; a
vector-summation motif, not neighbour coupling); $v\Delta$ is local but weak. The local positive
feedback a WTA needs is absent.}
\label{fig:c3}
\end{figure}

\paragraph{The FC2 ring lacks local recurrent excitation.} A WTA additionally requires local positive
feedback. We find no local recurrent \emph{excitation} in the FC2 wiring (Fig.~\ref{fig:c3}):
\begin{itemize}[nosep]
  \item $\text{FC2}\!\to\!h\Delta\!\to\!\text{FC2}$ recurrence (sign undetermined; measured as
        connectivity geometry, not transmitter) is \emph{anti-local}: it couples bearings
        $\sim$180$^\circ$ apart (a vector-summation motif; disynaptic weight $\sim$38 near, dipping
        mid-range, peaking at $\sim$95 antipodally, far/near ratio $2.5$), not neighbours.
  \item $v\Delta$-mediated recurrence is spatially local but weak ($5\%$ of $h\Delta$) and of
        undetermined sign.
  \item Direct $\text{FC2}\!\to\!\text{FC2}$ connectivity is negligible.
\end{itemize}
This geometry, global and anti-local inhibition, with no local recurrent excitation, is not the
wiring of a winner-take-all. Whether the circuit can nonetheless latch a choice we settle dynamically
in Sec.~\ref{sec:res-nowta} (Fig.~\ref{fig:c4c5}).

\paragraph{What the three routes sum to.} On the common disynaptic scale, the total
FC2$\to$FC2 inhibitory pathway is a \emph{large distance-uniform component plus a small anti-local
one}: \fbfa{} supplies $\sim$95\% of the total inhibitory mass but is flat across bearing (far/near
$1.0$, modulation $5\%$), contributing no distance-dependence; the $h\Delta$ route supplies only
$\sim$5\% of the mass yet carries essentially \emph{all} of the distance-dependence ($\sim$85\% of the
total's cross-bin variance, a covariance-adjusted attribution; far/near $2.5$ in isolation). So the connectome predicts inhibition of the
sign Mussells Pires reported, increasing from near to far bearings, but as a modest anti-local rise
on a uniform floor rather than a steep gradient; and because a normalizer subtracts the uniform \fbfa{}
term (Sec.~\ref{sec:res-normalizer}), it is the $h\Delta$ component that survives functionally. Two
caveats: synapse count is a structural proxy for functional inhibition, and we compare shape against
Mussells Pires' reported sign, not their digitized curve (which we do not have), a decomposition,
not a numerical fit.

\begin{figure}[t]\centering
\includegraphics[width=0.82\textwidth]{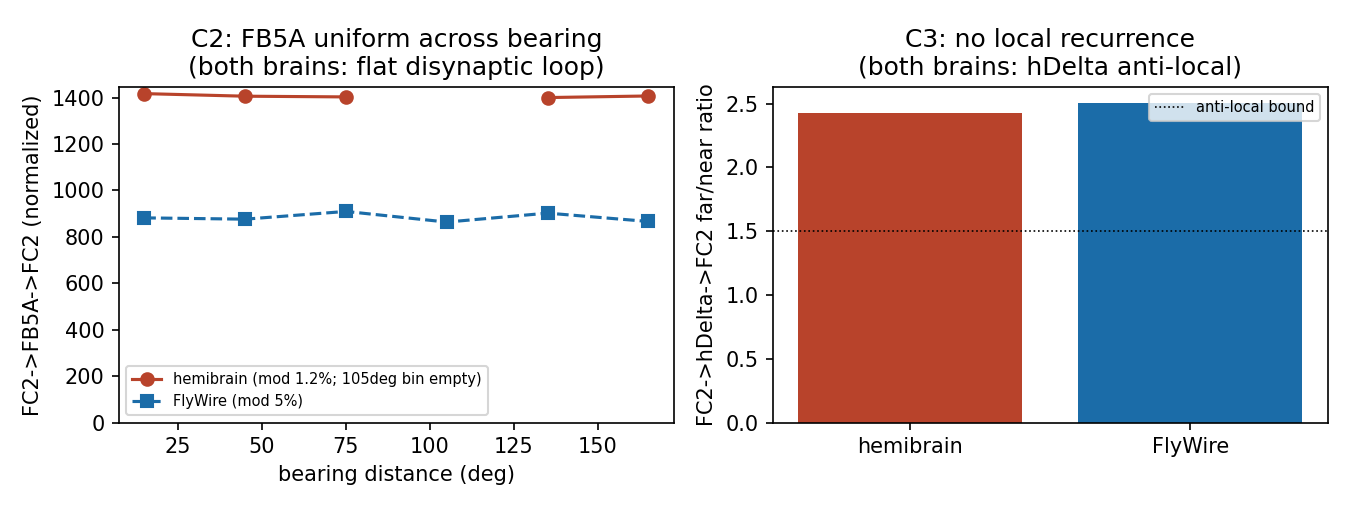}
\caption{\textbf{Two brains agree.} The FB5A-uniformity (left) and the $h\Delta$ anti-local
recurrence (right) reproduce in both the FlyWire and hemibrain connectomes
(\texttt{analyze\_fc2\_scoping\_hemibrain.py}). The dotted line on the right marks the anti-local
threshold (far/near ratio $1.5$). The hemibrain's $105^\circ$ bearing bin is empty
(partial-volume dataset), hence the gap in that curve.}
\label{fig:w3}
\end{figure}

\paragraph{The structure replicates in a second connectome.} In the independently reconstructed
hemibrain (Fig.~\ref{fig:w3}), both structural facts hold: \fbfa{} reaches all FC2 (88/88), the
disynaptic loop is flat (modulation $1.2\%$), direct $\text{FC2}\!\to\!\text{FC2}$ is negligible, and
$\text{FC2}\!\to\!h\Delta\!\to\!\text{FC2}$ is anti-local with far/near ratio $2.43$, matching
FlyWire's $2.50$ (at full precision $2.429$ hemibrain vs.\ $2.504$ FlyWire, a $3\%$ difference), across
two brains reconstructed by different groups. (The hemibrain is a partial volume, so the raw per-cell \fbfa{} weight carries a
completeness gradient that tracks per-column total synapses at $r{=}0.93$; the bearing-uniformity
check therefore uses the completeness-normalized weight (which drops the residual bearing correlation
from a raw $0.35$ to $0.145$), and the gradient is reported, not hidden.)

\subsection{No within-FC2 ring-attractor winner-take-all}
\label{sec:res-nowta}

\begin{figure}[t]\centering
\includegraphics[width=0.66\textwidth]{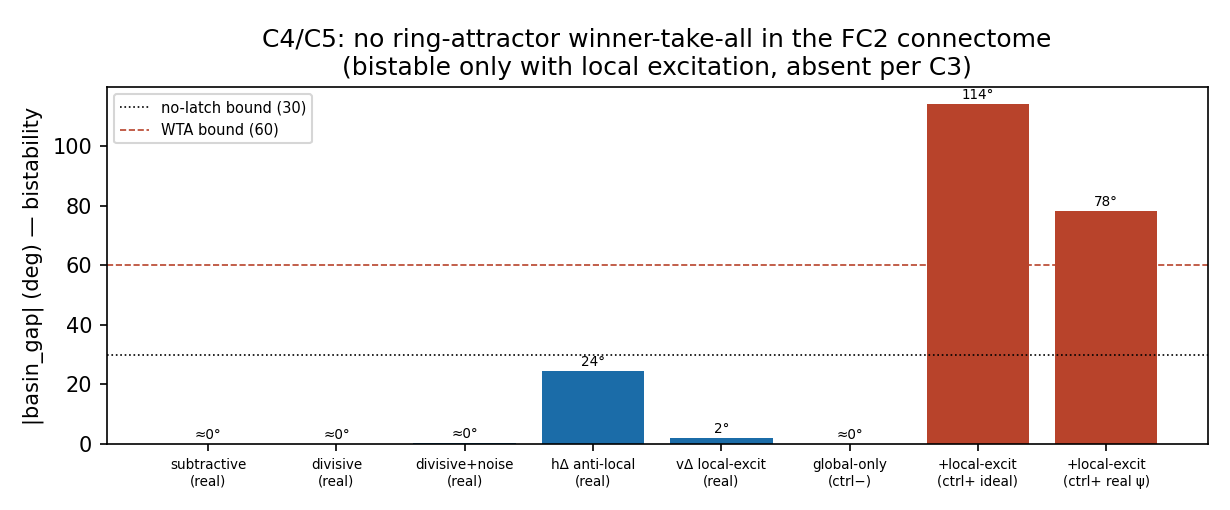}
\caption{\textbf{C4/C5, no bistability, hence no WTA.} Seeded-basin test: the pure global-inhibition
families (subtractive, divisive, divisive$+$noise), the one local route $v\Delta$ \emph{run as local
excitation} ($2^\circ$), and a global-only control ring have
$|\text{basin gap}|\approx 0^\circ$ (one input-determined fixed point). The $h\Delta$ anti-local
mutual-inhibition family sits at $24^\circ$ at the reference gain, below the $30^\circ$ no-latch
bound but the one gain-contingent case (it crosses that bound above gain $\approx\!6$, peaking at
$38.4^\circ$, a third of the $114^\circ$ positive control; see text). Only the
positive controls, local excitation added to the idealized ring ($114^\circ$) \emph{and} to the
real irregular bearings ($78^\circ$), are both bistable, confirming the test detects a WTA when one
exists, on the real geometry too (so the negative results are not an artifact of irregular sampling
suppressing detection), and identifying local excitation, which the real FC2 ring structurally lacks
(C3), as the ingredient that would be needed (\texttt{prove\_fc2\_no\_wta.py}; the committed leaky
integrate-and-fire run, basin gap $0^\circ$, is reported in the text and \texttt{analyze\_fc2\_lif.py}
but not drawn as a bar here).}
\label{fig:c4c5}
\end{figure}

We asked whether \emph{any} plausible dynamical model of the FC2$+$\fbfa{} circuit, parameterized by
the real connectome, collapses competing goals to a single winner (Sec.~\ref{sec:m-bistab}).
None commits. Divisive normalization \emph{shifts} the bump toward the stronger candidate (selection
by salience) but does \emph{not} collapse to a single winner, not even at 20\% amplitude
asymmetry. A seeded-basin test finds no bistability in any connectome-parameterized family (basin gap
$\approx 24^\circ$ or less at the reference gains, below the $30^\circ$ no-latch bound and far short of
the $114^\circ$ bistable control): no family latches within the no-latch bound ($h\Delta$, at
$24^\circ$, is closest). The spiking-proxy also fails,
as does a committed LIF spiking network (basin gap $0^\circ$); absent local recurrent excitation there
is no ring-attractor for spike timing to latch into. The structural core, the absence of the local
recurrent excitation a ring-attractor WTA needs (Sec.~\ref{sec:res-decomp}), holds across
formalisms: \textbf{there is no within-FC2 ring-attractor winner-take-all in the connectome}
(Fig.~\ref{fig:c4c5}); the $h\Delta$ mutual-inhibition route is the one gain-contingent exception,
confronted below.

\begin{figure}[t]\centering
\includegraphics[width=0.92\textwidth]{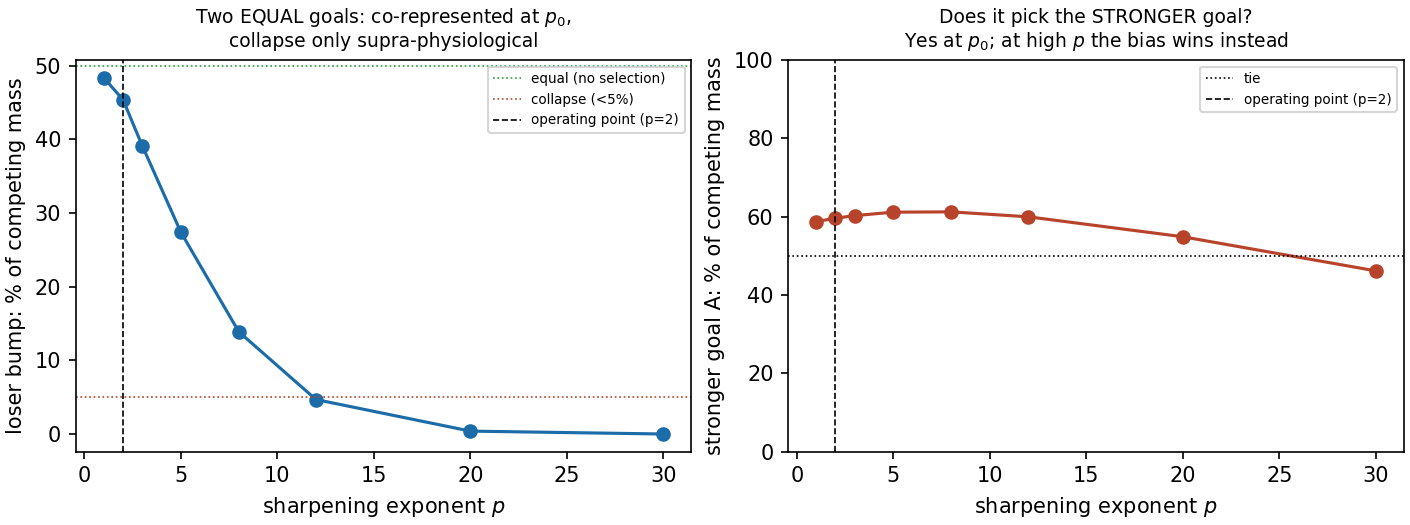}
\caption{\textbf{No winner-take-all via the feedforward sharpening term.} Left: with two \emph{equal}
competing goals, the loser bump keeps $\sim$half the mass at the operating point ($p{=}2$) and
collapses only supra-physiologically ($p\!\gtrsim\!12$); that collapse is init-independent (a fixed
bias, not a bistable choice). Right: the sharpening picks the \emph{stronger} goal at $p_0$ but at
high $p$ the fixed anatomical bias wins instead (stronger input drops below $50\%$), so it never
functions as a goal-selector (\texttt{analyze\_fc2\_feedforward.py}).}
\label{fig:ff}
\end{figure}

\paragraph{No feedforward max-selection either.} A bistability test rules out a \emph{recurrent} WTA,
but a divisive normalization with a high enough exponent can approach a hard max-selector without any
recurrence. Sweeping the sharpening exponent $p$ (Fig.~\ref{fig:ff}), at the task-agnostic operating
point ($p{=}2$) the circuit \emph{co-represents} competitors: two equal goals leave both bumps
($45\%/55\%$), and the stronger of two unequal goals leads with the weaker retained ($60\%/40\%$).
Collapse to a single column appears only at $p\!\gtrsim\!12$, six-fold above the operating point, 
and even then it is \emph{not} a winner-take-all: it is init-independent (seed-spread $0.00\%$, as
expected for an input-clamped map), so it amplifies a fixed anatomical bias rather than a
history-dependent choice; and it does \emph{not} track the goal, at high $p$ the \emph{stronger}
input loses ($46\%$ at $p{=}30$). The feedforward route yields no goal-WTA at any exponent, confirming
the bistability result from an independent axis.

\paragraph{Reconciling ``single-bump inhibition'' with ``no winner-take-all''.} This is not a
contradiction of \citet{mussellspires2024}: their distant inhibition is exactly what our decomposition
attributes to the uniform \fbfa{} and anti-local $h\Delta$ terms. But this geometry, global plus
anti-local, no local excitation, is not the machinery of a WTA. The FC2 circuit has a single
input-determined fixed point; it sharpens and normalizes the input but cannot latch a self-sustaining
choice.

\paragraph{The $h\Delta$ route, confronted.} A skeptic will press the $h\Delta$ family, and rightly: a
two-goal winner-take-all needs only mutual inhibition between the competitors, not a continuous ring
attractor, and $h\Delta$ supplies exactly that between antipodal columns. Worse, the behaviour we
anchor to, two goals $180^\circ$ apart yield a committed choice, not an average
(Sec.~\ref{sec:res-behaviour}), is the \emph{antipodal} configuration where $h\Delta$ coupling
peaks. So we state the alternative plainly: the FC2$\leftrightarrow h\Delta$ loop is the leading
candidate for an \emph{in-FC2} two-goal selector, and our result does not exclude it, it bounds it.
At the connectome-scaled reference gain its basin gap is $24^\circ$ (below the $30^\circ$ no-latch
bound); sweeping the anatomically-unconstrained scalar gain, it crosses that bound above
gain\,$\approx$\,6 ($32^\circ$) and peaks at $38.4^\circ$ (gain 8), but \emph{never} reaches
the $60^\circ$ WTA bound at any gain we tested (and is non-monotonic beyond, $32^\circ$ at gain 10).
So the honest statement is intermediate: $h\Delta$ shows a partial, gain-dependent seed-dependence but
does not reach full winner-take-all in the tested range, and the $30^\circ$ bound is a chosen
threshold, not a measured one. What does \emph{not} depend on any of this: the pure global-inhibition
families do not latch at any gain, and the one \emph{local} route, $v\Delta$, does not latch
\emph{even run as local excitation and swept over the same gain range as $h\Delta$} (basin gap
$\le 2^\circ$ throughout, never crossing the no-latch bound, unlike $h\Delta$), so ``no local
excitation'' is a gain-robust tested result, not a sign assumption or a claim only about the reference
gain. Whether the biological $h\Delta$ gain sits in the latching regime is
the single measurement that would decide between our normalization account and an $h\Delta$ selector,
and it is exactly what silencing or scaling $h\Delta$ during two-goal imaging would test. We flag this
as the primary open alternative, not a closed case.

\subsection{\fbfa{} as a uniform normalizer, distributedly sourced}
\label{sec:res-normalizer}

If \fbfa{} is not a spatial selector, the natural reading of a uniform inhibitory pool is
\textbf{divisive normalization} (a soft $k$-WTA): it scales the population by its total activity, so
the strongest drive survives best and a single clean bump is favoured, without ever being
\emph{forced}. This is the motif the mushroom body uses: the APL neuron provides feedback inhibition
to the Kenyon cells, producing a sparse, normalized code \citep{lin2014, flyhash2017}. Computationally,
and \emph{if \fbfa{} is inhibitory}, \fbfa{}:FC2 and APL:KC play the same role. This is an
instance of a canonical motif, not a new principle: the contribution is the cell-type assignment. We
use ``global'' in the computational sense; anatomically even APL inhibition is spatially graded
\citep{amin2020}, so the shared identity is the \emph{normalization computation}.

\begin{figure}[t]\centering
\includegraphics[width=0.92\textwidth]{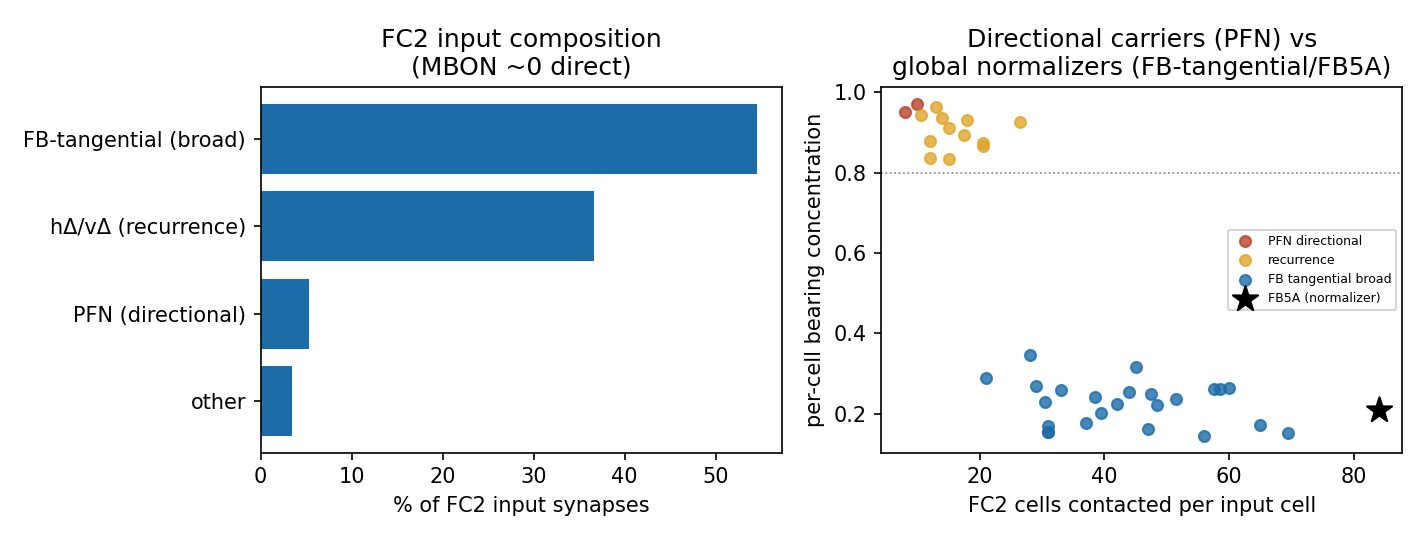}
\caption{\textbf{FC2's inputs: composition and per-cell geometry.} Left: input composition
(FB-tangential dominant; MBON $\approx$0 direct). Right: per-cell targeting, PFN cells are narrow
(few FC2 targets), \fbfa/FB-tangential cells are broad (bearing-uniform), the geometry a normalizer
would use (\texttt{trace\_fc2\_inputs.py}). The direct PFN$\to$FC2 edge is minor; the canonical PFN
route into FC2 is via $h\Delta$.}
\label{fig:w4}
\end{figure}

If FC2 does not itself select, the choice is shaped by its inputs, consistent with the fan-shaped
body as a context-dependent action-selection centre \citep{hulse2021} integrating distributed drives
into a goal \citep{matheson2022}. Our measurements localize the selector \emph{outside} the FC2 ring.
Tracing FC2's inputs (Fig.~\ref{fig:w4}): FC2 is fed (100\% typed) by FB-tangential neurons (55\%; a
broad family including \fbfa), $h\Delta$/$v\Delta$ interneurons (37\%), and PFN neurons (5\%; $\sim$3\%
other); there is \emph{no} direct input from mushroom-body output neurons (MBONs contribute
$\approx$0\% directly, so learned valence reaches FC2 indirectly). At single-cell resolution these
split into two geometries: each PFN cell targets a \emph{narrow} bearing band ($\sim$10 of 85 FC2,
concentration $0.96$), while each \fbfa{} cell blankets nearly the whole population (median 84/85 per
cell; all four collectively 85/85, concentration $0.21$).

\begin{figure}[t]\centering
\includegraphics[width=0.92\textwidth]{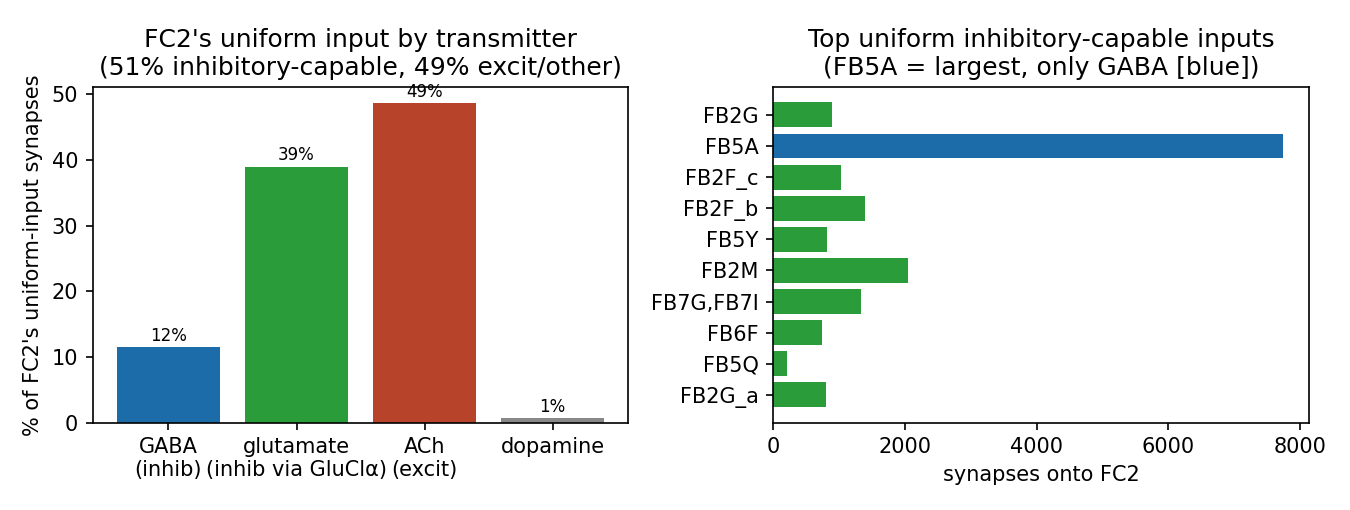}
\caption{\textbf{Who supplies FC2's uniform input.} Left: FC2's uniform-input synapse mass by
predicted transmitter, $\sim$51\% inhibitory-capable (GABA $+$ glutamate/GluCl$\alpha$), $\sim$49\%
cholinergic/other (excitatory). Right: the top uniform inhibitory-capable inputs by synapse count, \fbfa{}
(blue) is the largest and the only GABA-predicted one, atop a broad glutamatergic FB-tangential family
(green) (\texttt{analyze\_fc2\_uniform\_inhibitors.py}).}
\label{fig:cui}
\end{figure}

\paragraph{The uniform-inhibition substrate is distributed, not a one-cell bet.} Because \fbfa's
inhibitory identity is a low-confidence prediction (Sec.~\ref{sec:lim}), we asked which cells supply
FC2's \emph{uniform} drive, and of what transmitter (Fig.~\ref{fig:cui}). \fbfa{} is the
\emph{largest} single uniform input (7{,}741 synapses, 85/85) and the \emph{only} GABA-predicted one,
so naming it is not arbitrary, but only $\sim$12\% of the uniform-input mass. The rest is a
large family of FB-tangential cells that is predominantly \emph{glutamatergic} by type count (40
types; $\sim$39\% of the uniform-input mass), a transmitter that can be inhibitory via GluCl$\alpha$
\citep{liuwilson2013}; GABA and glutamate together
make $\sim$51\% of the uniform-input mass inhibitory-\emph{capable}. This is a capability, not a
confirmed sign: the GABA leg has receptor plausibility from central-complex Rdl/GABA-B expression
\citep{enell2007}, but the larger glutamate leg requires GluCl$\alpha$ on FC2, which we have not
verified. Dorsal fan-shaped-body tangential cells can moreover \emph{co-transmit} glutamate and
acetylcholine \citep{jones2025}, so a predicted-glutamatergic cell is not guaranteed to be net
inhibitory. Two consequences: first, the \emph{inhibitory role is robust to \fbfa's specific
transmitter}, whether \fbfa{} is GABAergic or (like sibling FB5AA) glutamatergic, it is
inhibitory-capable (the normalization feedback loop we model is established only for \fbfa{} itself,
via its FC2$\to$\fbfa$\to$FC2 pooling; the broad family shows the uniform inhibitory \emph{input} is
not one cell's alone, not that they all form the loop). Second, a caveat the wiring forces: the uniform
input is $\sim$half inhibitory-capable and $\sim$half cholinergic (excitatory), so the connectome alone
cannot declare the \emph{net} uniform drive inhibitory. The earned claim is thus the \emph{computation}
plus a \emph{ranked candidate substrate}, with \fbfa{} its leading member; the experiment of
Sec.~\ref{sec:res-prediction} decides the single-cell assignment.

\subsection{Where the selector is: a connectome-nominated candidate}
\label{sec:res-selector}

\begin{table}[t]\centering\small
\caption{The four \fbfa/$h\Delta$ input-and-inhibition fractions most easily confused, with their
explicit denominators (each individually correct). Rows 1--2 use different denominators from each other
and from rows 3--4, so are not comparable; rows 3 and 4 share the same denominator (FC2 total input)
and are \emph{nested}, the $28\%$ directional $h\Delta$ is the columnar subset of the $37\%$
recurrent mass. Route fractions elsewhere (e.g.\ the MBON$\to$FB5AB valence path) are quoted as a
percentage of the target cell's input.}
\label{tab:denominators}
\begin{tabular}{@{}lll@{}}
\toprule
\textbf{Quantity} & \textbf{Of what (denominator)} & \textbf{Where} \\
\midrule
\fbfa{} $\sim$95\% & total disynaptic FC2$\to$FC2 \emph{inhibitory mass} & decomposition sum \\
\fbfa{} $\sim$12\% & FC2's \emph{uniform-input} synapse mass ($\ge$50\% coverage) & §\ref{sec:res-normalizer} \\
$h\Delta$/$v\Delta$ $\sim$37\% & FC2's \emph{total input} synapses (all recurrence) & §\ref{sec:res-normalizer} \\
directional $h\Delta$ $\sim$28\% & FC2's \emph{total input} synapses ($h\Delta$ only) & §\ref{sec:res-selector} \\
\bottomrule
\end{tabular}
\end{table}

\begin{figure}[t]\centering
\includegraphics[width=0.92\textwidth]{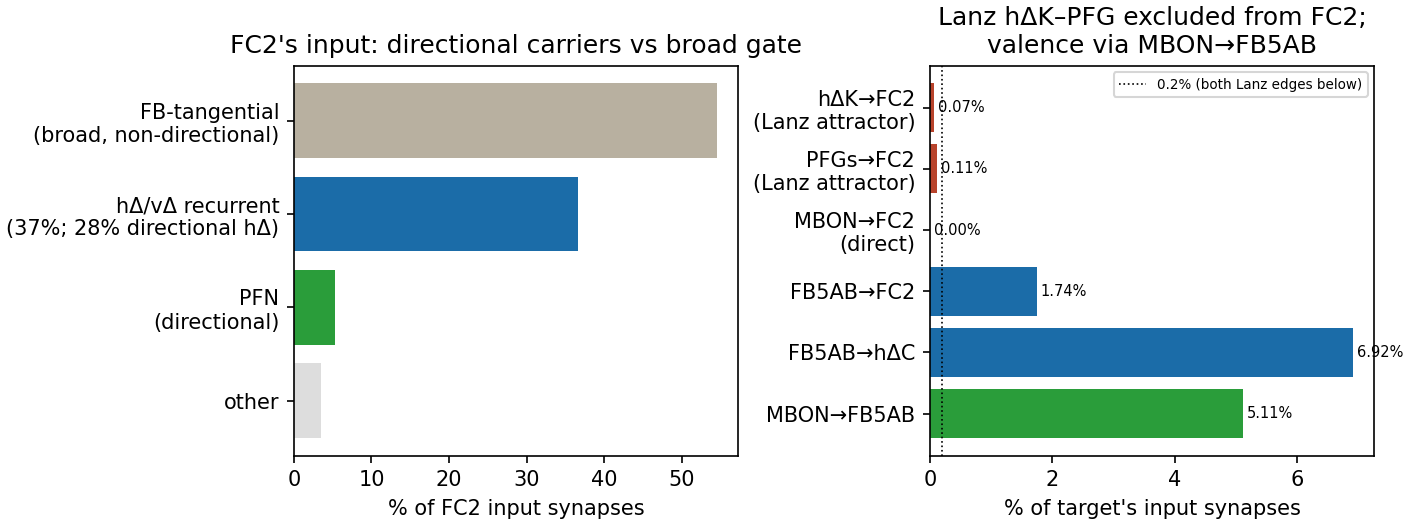}
\caption{\textbf{Where FC2's goal is set.} Left: FC2's input is dominated by a broad, non-directional
FB-tangential gate ($\sim$55\%); the directional carriers are the recurrent $h\Delta$ network and the
narrow PFN stream. Right: the Lanz $h\Delta$K--PFG persistent-goal attractor is excluded from FC2 (both
$<$0.2\%), while learned valence enters via MBON$\to$FB5AB$\to\{h\Delta$C, FC2$\}$
(\texttt{analyze\_fc2\_selector.py}).}
\label{fig:csel}
\end{figure}

Having placed goal selection \emph{outside} FC2, we ask the connectome to name the substrate that
holds it (Table~\ref{tab:denominators} lists the input fractions used here and their denominators).
FC2's input splits into a broad, non-directional FB-tangential stream ($\sim$55\%, per-cell
concentration $\sim$0.2, which cannot specify a bearing) and a structured columnar stream that can: the
recurrent $h\Delta$ network ($\sim$28\% of FC2 input, concentration $\sim$0.9) and a smaller
feedforward PFN vector input ($\sim$5\%, concentration $0.96$) (Fig.~\ref{fig:csel}). Of these, only
the $h\Delta$ network is at once \emph{directional}, \emph{high-drive}, and \emph{recurrent}
(within-class recurrence $\sim$10\%, the half-fan-shaped-body-offset motif of a vector memory
\citep{hulse2021}). It is also the only directional input that carries a learned signal: the sole route
from mushroom-body output into FC2's directional substrate is $\text{MBON}\!\to\!\text{FB5AB}\!\to\!
\{h\Delta\text{C},\text{FC2}\}$ (MBON$\to$FB5AB $5.1\%$; FB5AB$\to h\Delta$C $6.9\%$, its top named
input; FB5AB$\to$FC2 $1.7\%$; MBON$\to$FC2 direct $0\%$, each as a percentage of the target cell's
input), and FB5AB is a tangential neuron \emph{positioned to} gate an odour-and-wind goal heading in
$h\Delta$C \citep{matheson2022}. We rest the nomination on the \emph{anatomical}
MBON$\to$FB5AB$\to h\Delta$C connectivity above rather than on matching physiology, because the relevant
$h\Delta$C driver line is confounded: VT062617, used to record and perturb $h\Delta$C, was subsequently
reported to label the adjacent $h\Delta$K rather than $h\Delta$C \citep{matheson2024}, leaving those
functional results cell-type ambiguous (FB5AB itself is targeted by a separate line, 21D07). The
anatomical $h\Delta$K \emph{exclusion} from FC2 is unaffected by this line ambiguity. We therefore nominate the
\textbf{$h\Delta$ recurrent network, led by $h\Delta$C, with $h\Delta$J the largest single
$h\Delta$ input to FC2 ($6.1\%$), as the substrate that holds the committed goal FC2 reads}, set by
distributed directional and valence-gated drive.

\paragraph{A decisive exclusion.} The nomination separates FC2 from the best-modelled alternative. The
disinhibition-gated $h\Delta$K--PFG recurrent attractor \citep{lanz2025} is the strongest proposed
persistent-goal mechanism in the fan-shaped body, but its readout is PFL, not FC2. The connectome
adjudicates cleanly: $h\Delta$K and PFGs together supply \emph{$<$0.2\%} of FC2's input ($54$ and $88$
of $78{,}193$ synapses; $0.07\%$ and $0.11\%$), and $h\Delta$K projects to PFGs ($41\%$ of its output),
FB6A and ExR3, only $0.4\%$ to FC2. The $h\Delta$K--PFG attractor is thus a \emph{parallel,
PFL-facing} goal module, not the source of the FC2 bump. This yields a double dissociation: setting or
clamping the FC2 goal should require the $h\Delta$C/FB5AB pathway and be insensitive to $h\Delta$K/PFG
manipulation, whereas the PFL-side persistence of \citet{lanz2025} should show the opposite dependence.

\subsection{In-silico demonstration}
\label{sec:res-model}

\paragraph{Result 1, the single-goal bump shape is \fbfa-invariant.} That a single committed goal
yields one clean bump (resultant concentration $0.86$; $0.88$ raw) is how the operating point was
chosen (Sec.~\ref{sec:m-model}), so it is not itself a prediction. The load-bearing observation is that
silencing \fbfa{} leaves that concentration near-unchanged ($0.86\!\to\!0.85$): the bump's \emph{shape}
is set by the drive and the supralinear term, not by \fbfa{}, the premise the silencing prediction
(Result 2) builds on.

\begin{figure}[t]\centering
\includegraphics[width=0.82\textwidth]{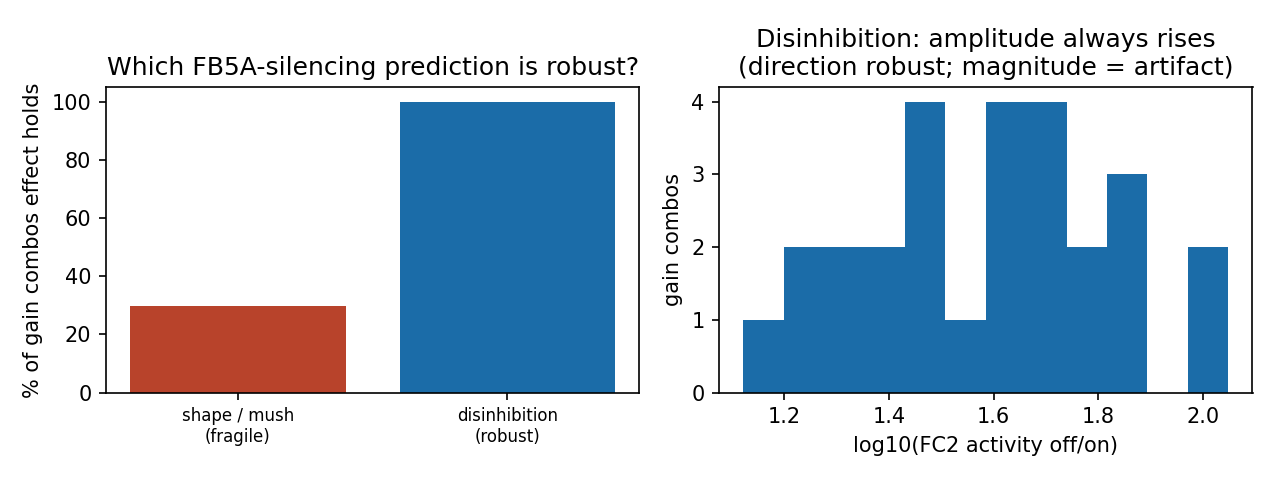}
\caption{\textbf{Which \fbfa{}-silencing effect is robust.} Left: across a $3\times3\times3$ gain
sweep, the shape/``mush'' effect holds in only $30\%$ of combinations (fragile), while the
disinhibition (amplitude-rise) effect holds in $100\%$. Right: the direction of the amplitude effect
is always the same (activity rises); its magnitude is a model artifact. The paper's prediction is the
robust one (\texttt{analyze\_fc2\_prediction.py}).}
\label{fig:u1}
\end{figure}

\paragraph{Result 2, what silencing \fbfa{} does, and what is robust.} We ran the silencing control
($g_{\text{inh}}=0$) and a $3\times3\times3$ gain sweep. Two candidate effects behave very differently
(Fig.~\ref{fig:u1}). (i) A \emph{shape} effect, competing candidates re-splitting the bump, 
appears at the reference gains but holds in only $30\%$ of the sweep: a threshold artifact we do
\emph{not} rely on (uniform inhibition cannot reshape a bump; the supralinear $[\cdot]^p$ term does the
sharpening). (ii) An \emph{amplitude} effect: silencing \fbfa{} disinhibits FC2. That activity
\emph{rises} is definitional (removing an additive denominator term); the non-trivial measurement is
that the bump's angular tuning is \emph{preserved} while activity rises (median $\Delta$concentration
$0.028$ across $100\%$ of the sweep). Disinhibition \emph{without reshaping} is the robust signature;
the \emph{magnitude} of the rise in this un-normalized limit is a model artifact and we make no
magnitude claim.

\subsection{Anchoring to fly choice behaviour}
\label{sec:res-behaviour}

The account must match what flies do when they choose. We anchor to four behavioural signatures, 
three measured in FC2, the fourth an analogy (flagged) \citep{mussellspires2024, toepfer2018}:
\begin{itemize}[nosep]
  \item The FC2 bump \textbf{tracks the goal} (bump--goal correlation $r{=}0.61$; the EPG heading bump
        tracks its cue more tightly, $0.88$).
  \item The FC2 bump--goal offset stays \textbf{clustered around zero under $\pm$90$^\circ$ virtual
        rotation} (offset distribution significantly directional; V-test $P{=}6.65\times10^{-4}$), 
        the bump tracks an allocentric goal, not a stimulus-locked response.
  \item Two goal bumps 180$^\circ$ apart yield two headings $\sim$166$^\circ$ apart, the fly
        \textbf{commits}, it does not average.
  \item Choice is \textbf{multi-stable}, stochastically switching between WTA-like and averaging-like
        regimes, consistent with a soft (normalizing) rather than hard competition. \emph{This last
        anchor is an analogy to multistable orientation behaviour under ambiguous motion
        \citep{toepfer2018}, hence a prediction, not a direct FC2 fit.}
\end{itemize}
These behaviours are consistent with a normalized, externally-set goal bump, and none requires a
within-FC2 winner-take-all, in line with Sec.~\ref{sec:res-nowta}. (The commitment signature is also
the antipodal configuration that presses the $h\Delta$ alternative; see Sec.~\ref{sec:res-nowta}.)

\subsection{A falsifiable prediction}
\label{sec:res-prediction}

The experiment is direct: silence \fbfa{} while imaging FC2 under two competing goals. It is a
\emph{single} experiment with a \emph{two-axis} readout forming a four-row discriminator (three
decisive outcomes plus a grey-zone intermediate) in which every outcome is informative (reading the
amplitude axis first, then the ratio):
\begin{center}
\begin{tabular}{@{}p{0.20\textwidth}p{0.34\textwidth}p{0.40\textwidth}@{}}
\toprule
\textbf{Amplitude} \emph{(total FC2 activity)} & \textbf{Ratio} \emph{(competing-column ratio)} & \textbf{Verdict} \\
\midrule
rises & preserved & global inhibitory \fbfa{}, \textbf{normalizer or uniform bystander} (same signature; C2 favours normalizer) \\
rises & changes which column dominates & a within-FC2 \textbf{local selector} (would overturn C2) \\
flat / falls & any & \fbfa{} is \textbf{not inhibitory} (transmitter prediction wrong, cf.\ \citealp{jones2025}) \\
rises & distorted but not flipped & \textbf{partially non-uniform} inhibition (grey zone) \\
\bottomrule
\end{tabular}
\end{center}
The \textbf{ratio} axis is the \emph{locality} test; the \textbf{amplitude} axis is the
\emph{mechanism} test.

\begin{figure}[t]\centering
\includegraphics[width=0.92\textwidth]{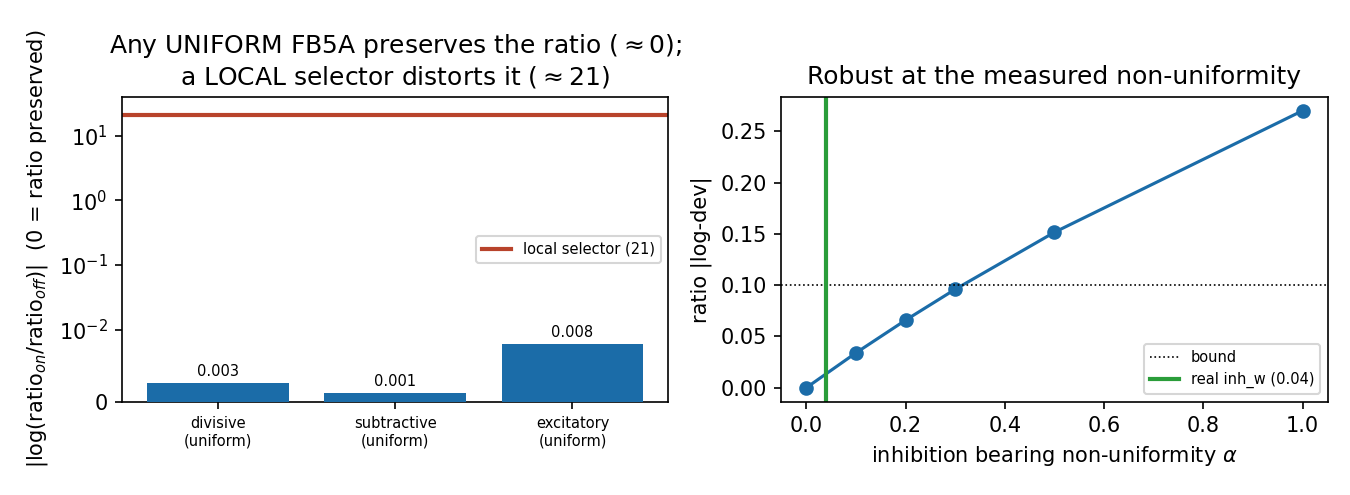}
\caption{\textbf{The two-cue test distinguishes a uniform \fbfa{} from a local selector.} Left: under
\fbfa{} silencing, the competing-column ratio is preserved for every \emph{uniform} functional form
(divisive/subtractive/excitatory, $|\log$-dev$|\!<\!0.01$) but distorted by a local-excitation selector
control ($\approx$21). Right: the ratio survives inhibition non-uniformity up to $\alpha\!\approx\!0.3$;
the measured \fbfa{} non-uniformity ($0.04$) sits well inside the preserved regime
(\texttt{analyze\_fc2\_competition.py}).}
\label{fig:u6}
\end{figure}

\paragraph{Locality axis, competition structure is preserved.}
\begin{quote}
\textbf{Under two competing goals, silencing \fbfa{} preserves the \emph{relative} activation of the
competing FC2 columns (the same ratio, scaled up), whereas a within-FC2 local spatial selector would
change which column dominates.}
\end{quote}
Because a spatially-uniform effect changes both columns nearly in proportion, silencing it leaves the
column ratio nearly unchanged, and this holds across the three uniform functional forms we tested,
independent of \fbfa's sign (divisive, subtractive, excitatory-additive, the last the \citet{jones2025}
failure mode), all preserving the ratio to $|\log(\text{ratio}_{\text{on}}/\text{ratio}_{\text{off}})|
< 0.01$ versus orders of magnitude larger for a local selector ($|\log|\approx21$, floor-limited; a
$>$2000-fold gap in the log-deviation metric) (Fig.~\ref{fig:u6}). A preserved ratio is the
\emph{expected} outcome given C2 (a uniform \fbfa{} cannot flip the ratio), so this axis
\emph{confirms} the measured connectivity and would only surprise us if a within-FC2 local selector
existed after all; the mechanism claim is carried by the amplitude axis. The ratio survives inhibition
non-uniformity up to $\alpha\!\approx\!0.3$, and the measured \fbfa{} non-uniformity is $0.04$, well
within margin. Preservation demonstrates \fbfa{} acts \emph{globally} (consistent with a normalizer,
but also with uniform gating or a bystander) and \emph{rules out} a within-FC2 local selector; it does
not by itself prove the normalizer role. Three wet-lab confounds bear on the measurement, the first a
genuine threat: because the fly \emph{commits} to one goal (Sec.~\ref{sec:res-behaviour}), there may be
no stable two-column ratio to read at all, and our workaround (read the transient pre-commitment
window, or impose sustained dual cues) is itself an assumption to validate; disinhibition-driven
saturation can compress the ratio; and silencing \fbfa{} leaves the anti-local $h\Delta$ term intact.

\paragraph{Amplitude axis, disinhibition without reshaping (the mechanism test).}
This axis discriminates an \emph{inhibitory} \fbfa{} from a non-inhibitory one, the distinction the
locality axis cannot make (it does not separate a normalizer from a uniform inhibitory bystander; C2
does that): silencing an inhibitory, normalizing \fbfa{} should \emph{disinhibit} FC2, raising overall
bump activity while preserving angular tuning, rather than reshaping the goal (Fig.~\ref{fig:u1}). We
do not lead the whole prediction with this axis because, unlike the locality axis, it \emph{depends on
\fbfa{} being inhibitory}; a flat or falling amplitude would instead be evidence that the predicted
GABAergic identity is wrong (the third table row, the \citet{jones2025} scenario). That contingency is
a feature of the joint readout: each decisive outcome tells us something.

% ============================================================================
\section{Discussion}
\label{sec:discussion}

\paragraph{What this paper establishes.} (i) A cell-type-resolved \textbf{decomposition} of the distant
feedback inhibition reported by \citet{mussellspires2024}: a uniform \fbfa{} term, an anti-local
$h\Delta$ term, a negligible direct term. (ii) A tested \textbf{constraint}: this geometry cannot host
a within-FC2 ring-attractor winner-take-all, verified across model families with a bistability test,
the wiring replicating in a second connectome; the $h\Delta$ mutual-inhibition route (partial
gain-dependent seed-dependence, never reaching full WTA) is flagged, not hidden. (iii) A \textbf{connectome-nominated upstream substrate} for the
goal FC2 reads (the $h\Delta$C-led recurrent network, valence-gated via FB5AB) and a \textbf{decisive
exclusion}, the Lanz $h\Delta$K--PFG attractor supplies $<$0.2\% of FC2's input, so it is
PFL-facing, not FC2's source. (iv) A \textbf{falsifiable prediction} that turns on the measured
uniformity, \fbfa{} silencing preserves competition structure (a global \fbfa) rather than changing
which column dominates (a local selector), plus a disinhibition corollary.

\paragraph{Interpretation.} Together these place FC2 as a \emph{normalizer of an externally-set goal},
not its selector: the goal is written into an upstream $h\Delta$ substrate (directional, recurrent,
valence-gated) and read out by PFL, while \fbfa{}-type global inhibition keeps the FC2 representation a
single clean bump. This reconciles the ``single-bump'' inhibition of \citet{mussellspires2024} with the
absence of a ring attractor, and separates FC2's goal from the parallel $h\Delta$K--PFG attractor
\citep{lanz2025}. The novel, transmitter-independent pieces are the decomposition, the no-WTA
constraint, and the exclusion; the normalizer role and the substrate nomination are the offered,
falsifiable interpretations.

% ============================================================================
\section{Limitations}
\label{sec:lim}

\paragraph{Prediction, not confirmation.} No recordings of the FC2 selection mechanism exist, and the
underlying distant-inhibition \emph{phenomenon} is not ours, it was reported by
\citet{mussellspires2024}; our contribution is its connectome decomposition and computational
character.

\paragraph{The transmitter is a low-confidence prediction.} FlyWire's classifier labels \fbfa{}
GABAergic at only $\sim$0.79 confidence, with no experimental verification, and the evidence is
unfavourable: one sibling type is experimentally \emph{GABA-negative} (FB5AB, cholinergic
\citep{matheson2022}), the other predicted-glutamatergic (FB5AA), and FB-tangential neurons are predominantly
glutamatergic, of the experimentally-typed FB-tangential cells in FlyWire, $14$ are GABAergic versus
$371$ GABA-negative. A direct precedent underlines the risk \citep{jones2025}. We therefore treat
\fbfa's identity as \emph{likely non-GABAergic on balance}, and rest the interpretation on the weaker
property of being \emph{inhibitory}: reachable via GABA (Rdl/GABA-B in the central complex \citep{enell2007}) or
glutamate via GluCl$\alpha$ \citep{liuwilson2013}. Neither route is verified on FC2, the GABA route
needs \fbfa{} GABAergic (judged unlikely), the glutamate route needs GluCl$\alpha$ on FC2 (unconfirmed),
so the inhibitory interpretation is a genuine open question; it fails outright only if \fbfa{} is
cholinergic (though the sign-independent locality prediction still survives even then), and is unconfirmed if \fbfa{} is
glutamatergic and FC2 lacks GluCl$\alpha$. The \emph{structural} results use only connectivity and
hold regardless of transmitter; the normalizer interpretation we present as a distributed,
evidence-ranked substrate (\fbfa{} its largest and only GABA-predicted member), not a bet on one cell.
(\fbfa{} here is the cell type labelled FB5A, distinct from FB5AA/FB5AB.) The \emph{divisive}
(shunting) form is a further declared modelling assumption, and the multi-stability anchor is an
analogy \citep{toepfer2018}, not a direct FC2 fit.

\paragraph{The $h\Delta$ selector alternative.} As stated in Sec.~\ref{sec:res-nowta}, an
$h\Delta$-mediated in-FC2 two-goal selector is not excluded, only bounded: it does not latch at the
reference gain but its anatomically-unconstrained gain is unmeasured, and the commitment behaviour
coincides with the antipodal regime where $h\Delta$ peaks. Deciding this needs $h\Delta$ physiology.

\paragraph{Method limits.} (a) The FC2 preferred bearings $\psi_i$ are inferred from the FC2/PFL
columnar fingerprint (not from \fbfa{} or $h\Delta$, so those tests are not circular); the hemibrain
uses the real anatomical column index, an independent axis. (b) Irregular real bearings can themselves resist clean latching; we checked this
directly with a positive control on the \emph{real} bearings (local excitation injected), which still
latches at $78^\circ$, above the $60^\circ$ WTA bound, so a real WTA would be detected, and the
negative results are not an artifact of the geometry (though the $78^\circ$ vs $114^\circ$ idealized
gap confirms the real sampling is somewhat conservative). (c) The distributed substrate is a structural argument; the in-silico model instantiates
only \fbfa. (d) The decomposition characterizes the routes' spatial signatures; we do not fit them to
reconstruct \citet{mussellspires2024}'s suppression-vs-distance curve. (e) \fbfa{} is only four cells,
small-population \emph{typing} is fragile; the 88/88 hemibrain replication mitigates the \emph{structural}
(coverage) fragility, not the $n{=}4$ transmitter prediction, which has no independent replication.

% ============================================================================
\section{Conclusion}
\label{sec:conclusion}

From a single connectome we decompose the fan-shaped-body goal circuit's feedback inhibition into a
uniform \fbfa{} term and an anti-local $h\Delta$ term, show the wiring cannot host a within-FC2
ring-attractor winner-take-all (tested across five model families, a committed spiking
run, and a feedforward sweep, with the wiring replicated in a second connectome), nominate the $h\Delta$C-led recurrent network as the upstream substrate
that holds the goal FC2 reads, and exclude the best-modelled alternative (the Lanz $h\Delta$K--PFG
attractor) as its source. FC2 normalizes an externally-set goal rather than selecting one. We interpret
\fbfa{} as the APL-type uniform normalizer and specify a two-axis silencing experiment that decides it,
alongside the one honestly open alternative, an $h\Delta$-mediated selector, whose gain is the
single measurement that would settle the account.

% ============================================================================
\section*{Code and data availability}
All wiring is sourced from the public FlyWire \citep{dorkenwald2024, schlegel2024} and hemibrain
\citep{scheffer2020, hulse2021} connectomes. All analysis code, the committed result caches, and the
test suite are available at \url{https://github.com/commonorigin/fc2-goal-circuit}; a versioned release
is archived at Zenodo (\url{https://doi.org/10.5281/zenodo.21252417}). Every figure and every reported number is
produced by a single re-runnable analysis script that writes a committed result cache; a committed test
suite checks each reported number against its cache, and every load-bearing number is traced to a
claims ledger in the repository. No parameter is fit to any downstream task.

\bibliographystyle{plainnat}
\bibliography{references}

\end{document}